\documentstyle[pra,aps,multicol,epsfig]{revtex}

\begin{document}
\title{Modulational instability of two-component Bose-Einstein\\
condensates in an optical lattice}
\author{Guang-Ri Jin$^{1}$, Chul Koo Kim$^{1}$, and Kyun Nahm$^{2}$}
\address{$^{1}$ Institute of Physics and Applied Physics, Yonsei University,
Seoul 120-749, Korea}
\address{$^{2}$ Department of Physics, Yonsei University, Wonju 220-710,
Korea}
\date{\today }
\maketitle

\begin{abstract}
We study modulational instability of two-component Bose-Einstein condensates
in an optical lattice, which is modelled as a coupled discrete nonlinear Schr%
\"{o}dinger equation. The excitation spectrum and the modulational
instability condition of the total system are presented analytically. In the
long-wavelength limit, our results agree with the homogeneous two-component
Bose-Einstein condensates case. The discreteness effects result in the
appearance of the modulational instability for the condensates in miscible
region. The numerical calculations confirm our analytical results and show
that the interspecies coupling can transfer the instability from one
component to another.\newline
\end{abstract}

\begin{multicols}{2}

\section{Introduction}
The modulational instability (MI) is a general phenomenon in
nonlinear wave equations and can occur in various physical
systems, such as fluid dynamics, plasma physics, and nonlinear
optics. Due to the interplay between nonlinearity and the
dispersive effects, a weak perturbation on a plane wave may induce
an exponential growth, and result in the carrier wave break up
into a train of localized waves \cite{Agrawal}. Recently, the MI
of Bose-Einstein condensates (BEC) has attracted much interest,
especially for the single-component BEC in an optical lattice
\cite{WNiu,Konotop}. In the superfluid regime, the BEC in an
optical lattice can be modeled by a discrete nonlinear
Schr\"{o}dinger equation \cite{Trombettoni01,Abd}. Following
Ref.\cite{Kivshar}, the MI of a BEC in a deep optical lattice has
been studied both theoretically
\cite{Trombettoni02,Smerzi02,Smerzi03,Rapti} and experimentally
\cite{Cataliotti,Fallani}.

Compared with the single-component BEC system, binary mixtures of
BECs have much richer structures \cite{TLHo}. In fact, the
excitation spectrum of a homogenous two-component BECs has been
studied in Refs. \cite{Meystre,Bashkin,Graham}. It was shown that
the MI depends strongly on the sign of $\Lambda_{12}^{2}-\Lambda
_{1}\Lambda _{2}$, where $\Lambda_{12}$ is the interspecies
interaction strength, $\Lambda _{1}$ and $\Lambda _{2}$ are the
intraspecies scattering strengthes, respectively. In Ref.
\cite{Rapti1}, the MI of the two-component BECs in a deep optical
lattice was studied with a Josephson-like coupling term. However,
their works were limited on a fixed perturbation wave number
(i.e., $q=\pi$). In addition, they did not compare their results
with that of the homogenous two-component BECs case
\cite{Meystre,Bashkin,Graham}.

In this paper, we study the MI of two-component BECs in an optical
lattice. The explicit expression of the excitation spectrum is
presented analytically. In order to outline the effects of the
discreteness, we compare the MI condition of the modulated plane
waves with that of the homogenous two-component BECs case, in
which the MI can occur only for $ \Lambda_{12}^{2}>\Lambda
_{1}\Lambda _{2}$, i.e., the BECs in phase separation region
\cite{Meystre,Bashkin,Graham}. Our results show that, the MI of
the two component BECs in an optical lattice depends not only on
the sign of $\Lambda _{12}^{2}-\Lambda _{1}\Lambda _{2}$, but also
on the wave number $k$ of the carrier waves. For
$\Lambda_{12}^{2}<\Lambda _{1}\Lambda _{2}$, the MI of the
miscible condensates can also take place when the condensates have
relatively large wave number $k>\pi /2$. The effect of the
interspecies coupling is investigated by numerical calculations.
Our results show that the coupling between the two condensates can
transfer the instability from one component to another.

\section{Theoretical model}
We consider a cloud of BEC atoms which have two internal states
labeled by $|1\rangle $ and $|2\rangle $ in an optical lattice.
When the heights of the interwell barriers are much higher than
the chemical potentials, the condensate wave functions can be
expressed as a sum of wave functions localized in each well of the
periodic potential $ \Psi _{\sigma }({\bf
r},t)=\sqrt{N}\sum_{j}\psi _{j,\sigma }(t)\phi _{\sigma }({\bf
r-r}_{j})$, where $N$ is the total number of condensate atoms, and
$\phi _{\sigma }({\bf r-r}_{j})$ is the spatial wave function
localized in the $j$th site. With the help of the tight-binding
approximation \cite{Trombettoni01}, the probability amplitudes
$\psi _{j,\sigma }$ obey a set of coupled DNLS equations ($\hbar
=1$):
\begin{eqnarray}
i\frac{\partial \psi _{j,\sigma }}{\partial t} &=&-K_{\sigma }(\psi
_{j-1,\sigma }+\psi _{j+1,\sigma })  \nonumber \\
&&+\left( \Lambda_{\sigma,\sigma}|\psi _{j,\sigma }|^{2}+\Lambda
_{\sigma \sigma ^{\prime }}|\psi _{j,\sigma ^{\prime
}}|^{2}\right) \psi _{j,\sigma }, \label{D2}
\end{eqnarray}%
where $K_{\sigma }$ and $\Lambda _{\sigma \sigma ^{\prime }}$ with
$\sigma =1,2$ describe the nearest-neighbor hopping term and the
on-site atomic collisions, respectively. The stationary solutions
of the DNLS equations (\ref{D2}) are plane waves $ \psi _{j,\sigma
}=\psi _{\sigma }^{(0)}e^{i(kj-\mu _{\sigma }t)}$ with the
chemical potentials of the two-component condensed atoms
\begin{equation}
\mu _{\sigma }=-2K_{\sigma }\cos (k)+\Lambda _{\sigma,\sigma
}|\psi _{\sigma }^{(0)}|^{2}+\Lambda _{\sigma \sigma ^{\prime
}}|\psi _{\sigma ^{\prime }}^{(0)}|^{2}.  \label{omega}
\end{equation}%
Here, the chemical potentials $\mu _{\sigma }$ are to be
determined by the normalization condition $\sum_{j,\sigma }|\psi
_{j,\sigma }|^{2}=1$. The terms $-2K_{\sigma }\cos (k)$ in Eq.
(\ref{omega}) are the kinetic energies aroused from tunnelling. It
should be noted that, we study the BECs system in the superfluid
regime with $(N/M)K_{\sigma}\gg \Lambda_{\sigma,\sigma'}$ ( $M$ is
the lattice number), and thus the dynamics can be well described
by the DNLS equations.

The excitation spectrum can be obtained by the stability analysis
of the plane waves. We introduce small perturbations to the
carrier wave as: $\psi _{j,\sigma }=\left[ \psi _{\sigma
}^{(0)}+\delta \psi _{j,\sigma }\right] e^{i(kj-\mu _{\sigma
}t)}$, where the small amplitude phonon modes are taken in the
traditional form $\delta \psi _{j,\sigma }=u_{\sigma
}e^{i(qj-\omega t)}+v_{\sigma }^{\ast }e^{-i(qj-\omega t)}$, with
the constants $u_{\sigma }$ and $v_{\sigma }$ to be determined.
Then we obtain the excitation spectra as
\begin{equation}
\omega _{q,\sigma }^{(\pm )} =2K\sin (k)\sin (q)\pm
\sqrt{\epsilon_{q}(\epsilon _{q}+\Delta _{\sigma })},
\label{spec2}
\end{equation}
where
\end{multicols}
\begin{widetext}
\begin{equation}
\Delta _{\sigma } =\Lambda _{1}(\psi _{1}^{(0)})^{2}+\Lambda
_{2}(\psi _{2}^{(0)})^{2}+(-1)^{\sigma }  \sqrt{\left[ \Lambda
_{1}(\psi _{1}^{(0)})^{2}-\Lambda _{2}(\psi _{2}^{(0)})^{2}\right]
^{2}+4\Lambda _{12}^{2}(\psi _{1}^{(0)}\psi _{2}^{(0)})^{2}}.
\label{vv}
\end{equation}
\end{widetext}
\begin{multicols}{2}
\noindent In the above derivations, we have assumed $\psi _{\sigma
}^{(0)}$ to be real, and taken $K_{1}=K_{2}=K$, then $\epsilon
_{q,1}=\epsilon _{q,2}=\epsilon _{q}\equiv 4K\cos (k)\sin
^{2}(q/2)$ \cite{Paraoanu}. In the absence of the interspecies
interactions ($\Lambda _{12}=0$), $\Delta _{\sigma }=2\Lambda
_{1}(\psi _{1}^{(0)})^{2}$ or $2\Lambda _{2}(\psi
_{2}^{(0)})^{2}$. The two-component BECs are fully decoupled, and
the spectra are the same with single-component case
\cite{Smerzi03,Paraoanu}. According to Refs. \cite
{Kivshar,Smerzi02}, the plane waves of the condensates are
modulationally unstable when the eigenfrequencies becomes
imaginary, i.e., $\epsilon _{q}(\epsilon _{q}+\Delta _{\sigma
})<0$. In this case, the system undergoes exponential growth with
the growth rate determined by the imaginary part of $ \omega
_{q,\sigma }^{(\pm )}$. From Eq. (\ref{vv}), we know that: (i) if
$ \Lambda _{12}^{2}<\Lambda _{1}\Lambda _{2}$, i.e., the case of
the two components being miscible, both $\Delta _{1}$ and $\Delta
_{2}\ $are positive; (ii) if $\Lambda _{12}^{2}>\Lambda
_{1}\Lambda _{2}$, then $\Delta _{1}<0$ and $\Delta _{2}>0$.
Within the second scenario, the condensates tend to separate
spatially, namely the phenomena of phase separation \cite
{Timm,AoChui}. The sign of $(\Lambda _{12}^{2}-\Lambda _{1}\Lambda
_{2})$ is of crucial important in discussing the stability of the
eigenmodes.

\section{Modulational instability}
In this section, we study the modulational instability by using
Eq. (\ref{spec2}). To understand the effect of the discreteness
clearly, we first study Eq. (\ref{spec2}) in the long-wavelength
limit \cite{Kivshar}.  Eq. (\ref{spec2}) in the long-wavelength
limit, i.e., $k\rightarrow 0$ and $q\rightarrow 0$, is reduced to:
\begin{equation}
\omega _{q,\sigma }^{(\pm )}=2Kqk\pm \sqrt{Kq^{2}(Kq^{2}+\Delta
_{\sigma })}, \label{LWLspec}
\end{equation}%
which agrees with the usual expression obtained from continuous
NLS equations \cite{Meystre,Bashkin,Graham} by replacing $Kq^{2}$
with $q^{2}/2m$. It was shown that only immiscible BECs (with
$\Lambda _{12}^{2}>\Lambda _{1}\Lambda _{2}$) exhibit the MI
\cite{Kasa}.

\subsection{The MI condition for the miscible BECs}
The MI condition for the two-component BECs are determined by the
condition: $\epsilon _{q}(\epsilon _{q}+\Delta _{\sigma })<0$. For
the miscible BECs with $\Lambda _{12}^{2}<\Lambda _{1}\Lambda
_{2}$, the MI occurs for $\cos (k)<0$ and $\epsilon _{q}+\Delta
_{\sigma }>0$. For simplicity, we take $\psi _{1}^{(0)}=\psi
_{2}^{(0)}=\psi _{0}$, then $\Delta _{\sigma }=\Omega _{\sigma
}\psi _{0}^{2}$, where
\begin{equation}
\Omega _{\sigma }=\Lambda _{1}+\Lambda _{2}+(-1)^{\sigma
}\sqrt{(\Lambda _{1}-\Lambda _{2})^{2}+4\Lambda _{12}^{2}}.
\end{equation}
Therefore, the miscible BECs may exhibits the MI when $\pi
/2<k<3\pi /2$ and $\psi _{0}^{2}>-\epsilon _{q}/\Omega _{\sigma
}$. More specially, both of the two-component with relative large
wave number ($k>\pi /2$) will be unstable for arbitrary $q$
provided that $\psi _{0}^{2}>\psi _{0,\text{cr}}^{2}=4K/\Omega
_{1}$. Below the critical amplitude, the MI region depends on the
perturbation's wave number $q$. In Fig. 1 (a), we consider the
case $\psi _{0}^2<\psi_{0,\text{cr}}^{2}$, where the atomic
collisions are taken as $a_{1}:a_{12}:a_{2}::1.007:1:1.01$
\cite{Burke}. We take $\Lambda_1=\Lambda$, then
$\Lambda_{12}=0.993\Lambda$ and $\Lambda_{2}=1.00298\Lambda$,
which satisfy the miscible criterion. Our results show that there
exist three different regions in the ($q$, $k$) plane. The dotted
line in Fig. 1(a) separates the unstable regime into a fully
unstable and a partially unstable, in which one component is
stable and the second one is unstable.

\subsection{The MI condition for the BECs in phase separation region}
To study the MI of the immiscible BECs, we take the atomic
collisions are taken as $a_{1}:a_{12}:a_{2}::1.03:1:0.97$
\cite{Hall}. Thus we have $\Lambda_{12}=0.9709\Lambda$ and
$\Lambda_{2}=0.9417\Lambda$, which meet the condition $\Lambda
_{12}^{2}>\Lambda _{1}\Lambda_{2}$. Unlike the previous miscible
case, the MI condition considered here depends on the wave number
$k$. We find that for $\cos k>0$ (i.e., $k<\pi /2$ or $k>3\pi
/2$), the MI condition is $\psi _{0}^{2}>\epsilon
_{q}/|\Omega_{1}|$. However, for $\pi/2<k<3\pi /2$, it becomes
$\psi _{0}^{2}>-\epsilon _{q}/\Omega_{2}$. Since
$|\Omega_{1}|\ll\Omega_{2}$, the immiscible BECs will be unstable
in the whole ($q$, $k$) plane provided that $\psi _{0}^{2}>\psi
_{0,\text{cr}}^{2}=4K/|\Omega_{1}|$. Below the critical amplitude,
the MI region in the ($q$, $k$) plane is shown in Fig. 1 (b). The
MI can take place for any wave number $k$.

\vskip -1.07cm
\begin{figure}[htbp]
\begin{center}
\epsfxsize=7cm\epsffile{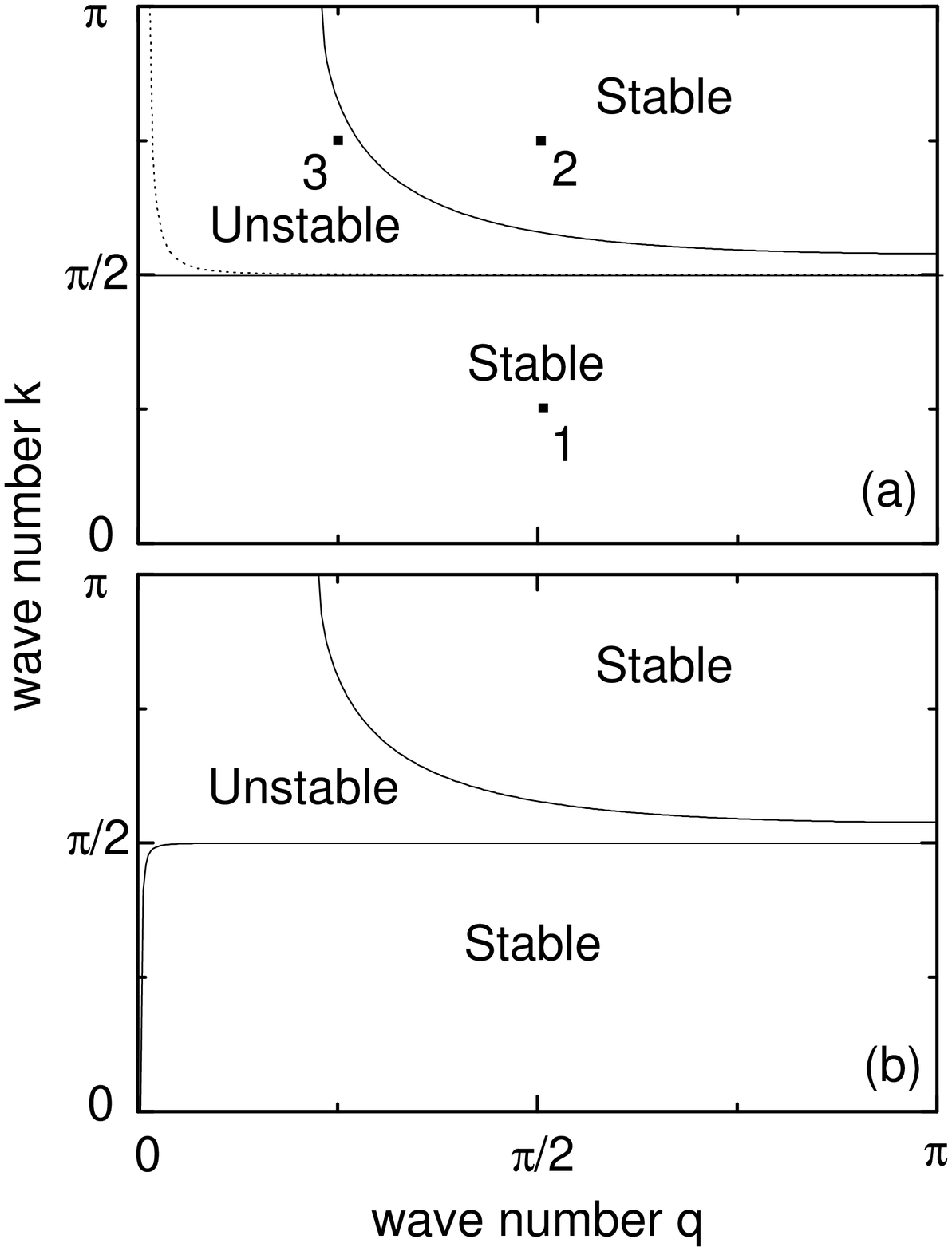}
\end{center}
\vskip -0.8cm \caption{Region of the MI in the ($q$, $k$) plane
for (a) miscible BEC $\Lambda_{12}=0.993\Lambda$ and
$\Lambda_{2}=1.00298\Lambda$, (b) immiscible BEC
$\Lambda_{12}=0.9709\Lambda$ and $\Lambda_{2}=0.9417\Lambda$.
Other parameters are taken as $K=1$, $\Lambda_1=\Lambda=100$, and
$\psi _{0}^{2}=1/(2M+1)<\psi _{0,\text{cr}}^{2}$.}
\end{figure}

\section{Numerical analysis}
The linear-stability analysis can determine the onset of MI,
however it does not yield any dynamical information beyond the
instability point. Therefore, we perform numerical simulations of
the DNLS equations. The initial conditions are two modulated plane
waves
\begin{equation}
\psi _{j,1}(0)=\psi _{j,2}(0)=[A+\alpha\cos (qj)] e^{ikj},
\end{equation}
in accordance with our previous discussions. The modulation
amplitudes $\alpha$ are assumed small compared with the background
amplitudes $\alpha=0.05A$. We consider the optical lattice with
the total number of the sites $M=400$. From the periodic boundary
condition, the wave numbers take the form of $k=2\pi l/M$ and
$q=2\pi s/M$, where we choose the unit lattice constant.

We first consider the miscible BEC case. We take the amplitude as
$A=1/\sqrt{2M+1}$ to insure the normalization condition. The wave
numbers are chosen as ($l=50$, $s=100$), ($l=150$, $s=100$), and
($l=150$, $s=50$), which correspond to the points labeled by 1, 2,
3 in Fig. 1(a). The first two points are stable, and the last one
is in an unstable regime with the growth rate being $0.1863$.
Density of the first component $|\psi_{j,1}(t)|^2$ is plotted in
Fig. 2. As expected from the analytical prediction, for the first
point ($k=\pi/4$, $q=\pi/2$), the modulated wave is stable.
However, for the second point ($k=3\pi/4$, $q=\pi/2$) as shown in
Fig. 2(b), the density increase sharply at the final stage. It is
not predicted by the linear stability analysis. The reason for
this phenomenon is the complex interactions between three
fundamental eigenmodes of wave numbers $k$, $k-q$, and $k+q$
contained in the initial state, and additional components of other
wave numbers \cite{Kivshar}. Fig. 2(c) shows the case that
$\omega_{q,1}$ is real and $\omega_{q,2}$ is imaginary, which
means that the component 1 is stable and the component 2 is
unstable. However, the instability is transferred from one
component to another due to the effects interspecies coupling,
which in turn leads to the MI of total system. The appearance of
spatially localized modes with large amplitude indicates the
modulational instability of the miscible BECs.

\vskip -1.07cm
\begin{figure}[htbp]
\begin{center}
\epsfxsize=7cm\epsffile{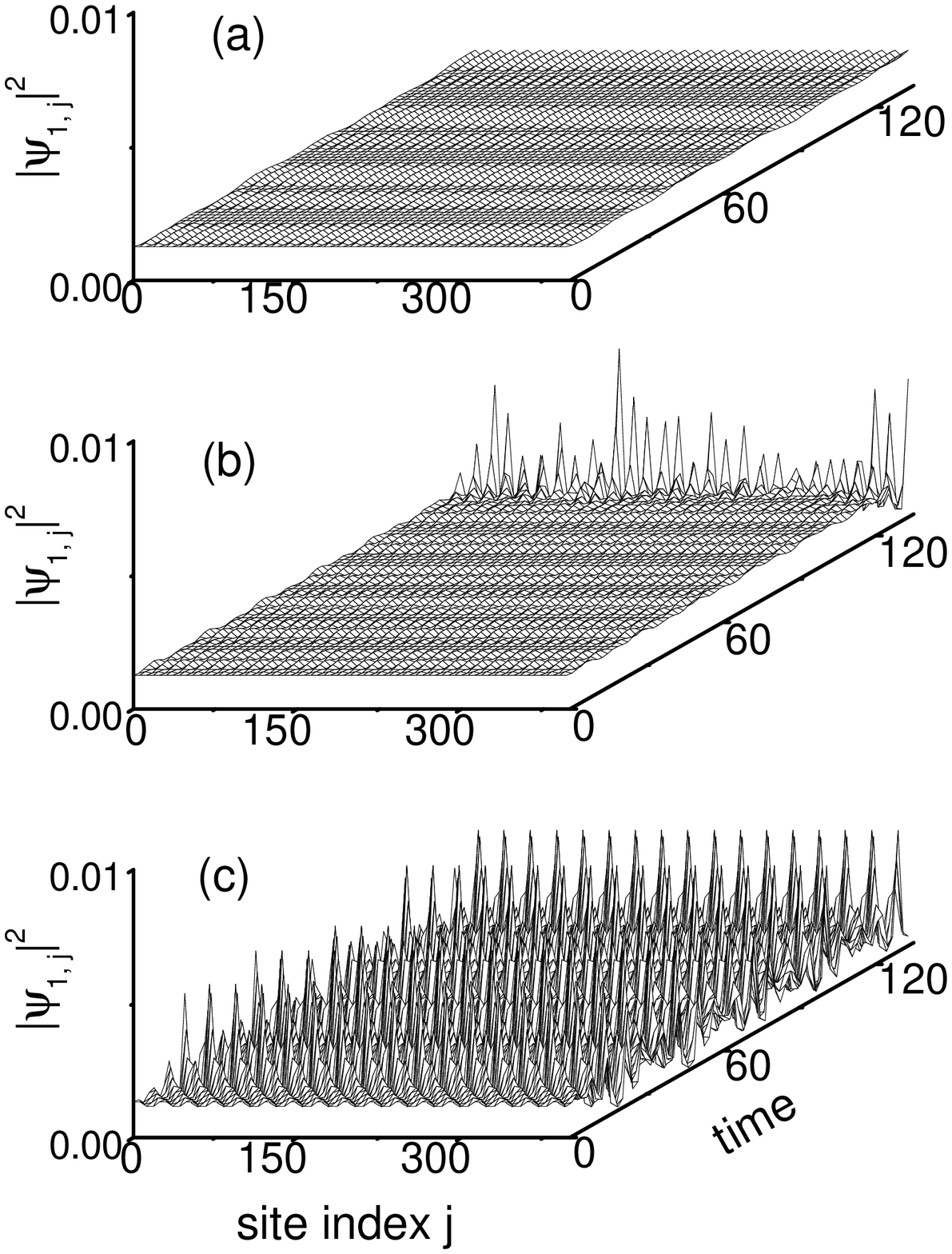}
\end{center}
\vskip -1.2cm \caption{Density of component 1 for the miscible
case ($\Lambda_{12}^2<\Lambda_1\Lambda_2$). Component 2 is similar
with component 1. (a) $k=\pi/4$, $q=\pi/2$, (b) $k=3\pi/4$,
$q=\pi/2$, and (c) $k=3\pi/4$, $q=\pi/4$. Other parameters are the
same as Fig. 1(a).}
\end{figure}

In Ref. \cite{Rapti1}, Rapti et al. calculated the spatiotemporal
dynamics of the two-component BECs with purely nonlinear coupling,
where the wave number $k$ is taken as $\pi/2$. According to the
linear-stability analysis, however, within this case both $\omega
_{q,1}$ and $\omega _{q,2}$ are real and equal to $2K\sin(q)$,
which means that there is no the MI. The appearance of the
large-amplitude excitations in their numerical simulation may
originate from the complex interactions between the eigenmodes
involved.

We further study the MI of the two-component BECs for the
immiscible case. As an example we calculate two points: ($l=150$,
$s=5$) and ($l=150$, $s=10$) in the unstable regime. Both points
have real $\omega_{q,1}$ and imaginary $\omega_{q,2}$. The MI
growth rates are $0.0458$ and $0.0902$, respectively. The
instability growth time for the second point is twice to that of
the first point with the second one. Our numerical results shown
in Fig. 3 confirm the analytical prediction. Other simulations in
the unstable regime show that the increase of $q$ in the monotonic
regime ($0<q<\pi$) can enhance the growth rate, i.e., the
instability growth time becomes more and more shorter with the
increase of $q$.

\vskip -1.07cm
\begin{figure}[htbp]
\begin{center}
\epsfxsize=7cm\epsffile{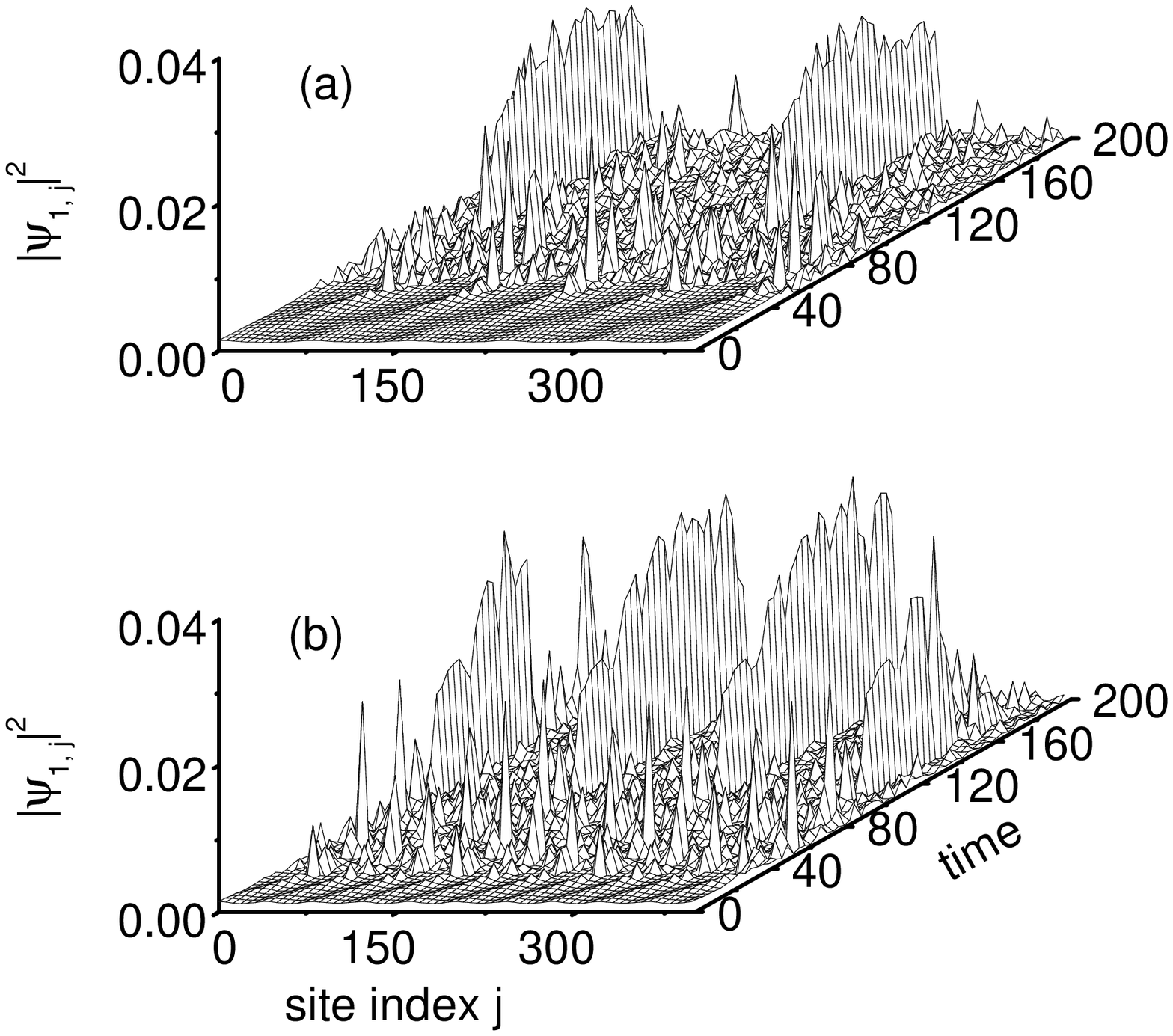}
\end{center}
\vskip -3.2cm \caption{Density of component 1 for the immiscible
case ($\Lambda_{12}^2>\Lambda_1\Lambda_2$). Component 2 is similar
with component 1. (a) $k=3\pi/4$, $q=\pi/40$, (b) $k=3\pi/4$,
$q=\pi/20$. Other parameters are the same as Fig. 1(b).}
\end{figure}

\section{Conclusions}
We have studied the modulational instability of two-component BEC
in the superfluid regime. The explicit expression of the
excitation spectrum is presented analytically. In the
long-wavelength limit, our results can recover previous results of
two-species BECs in homogeneous case. From the Bogoliubov
dispersion relation we studied the MI condition of the modulated
plane waves. The discreteness effect leads to the appearance of
the MI for the two-component BECs not only in the miscible region
but also in the phase separation region.

For the miscible condensates, the MI can occur for relatively
large wave number of the condensates $k>\pi/2$. For the smaller
$\psi _{0}^{2}$ compared with the critical density $\psi _{0,
\text{cr}}^{2}$, the MI condition depends also on the perturbation
wave number $q$. The MI conditions in the two-component BECs with
the phase separation can take place for any $k$. However, unlike
to the miscible case, only one of the species can be
modulationally unstable for any given $q$ and $k$. To confirm the
analytical results, we also performed numerical calculations. We
find that the interspecies coupling can transfer the instabilities
from one component to the other, which lead to the MI of the whole
system.

In summary, with the help of linear-stability analysis, we have
studied the stability of the modulated plane waves, which is
essential to predict the existence of nonlinear localized modes,
such as dark solitons, in the discrete nonlinear Schr\"{o}dinger
equation \cite{Kivshar94,Kevrekidis}.

This work is supported in part by the BK21 and by KOSEF through
Center for Strongly Correlated Materials Research, SNU. We would
like to express our sincere thanks to Dr. Guo-Hui Ding for helpful
discussions.

\end{multicols}


\begin{references}
\bibitem{Agrawal} G. P. Agrawal, {\it Nonlinear Fiber optics} (Academic
Press, San Diego, 2001).

\bibitem{WNiu} B. Wu and Q. Niu, Phys. Rev. A {\bf 64}, 061603(R) (2001).

\bibitem{Konotop} V. V. Konotop and M. Salerno, Phys. Rev. A {\bf 65},
021602(R) (2002); B. B. Baizakov, V. V. Konotop, and M. Salerno, J. Phys. B
{\bf 35}, 5105 (2002).

\bibitem{Trombettoni01} A. Trombettoni and A. Smerzi, Phys. Rev. Lett. {\bf %
86}, 2353 (2001).

\bibitem{Abd} F. K. Abdullaev {\it et al.}, Phys. Rev. A {\bf 64}, 043606
(2001)

\bibitem{Kivshar} Y. S. Kivshar and M. Peyrard, Phys. Rev. A {\bf 46}, 3198
(1992).

\bibitem{Trombettoni02} A. Trombettoni {\it et al.}, Phys. Rev. Lett. {\bf 88%
}, 173902 (2002).

\bibitem{Smerzi02} A. Smerzi {\it et al.}, Phys. Rev. Lett. {\bf 89}, 170402
(2002).

\bibitem{Smerzi03} A. Smerzi and A. Trombettoni, Phys. Rev. A {\bf 68},
023613 (2003).

\bibitem{Rapti} Z. Rapti {\it et al.}, J. Phys. B {\bf 37}, S257 (2004).

\bibitem{Cataliotti} F. S. Cataliotti {\it et al.}, New J. Phys. {\bf 5}, 71
(2003).

\bibitem{Fallani} L. Fallani {\it et al.}, Phys. Rev. Lett. {\bf 93}, 140406
(2004).

\bibitem{TLHo} T. L. Ho and V. B. Shenoy, Phys. Rev. Lett. {\bf 77}, 3276
(1996).

\bibitem{Meystre} E. V. Goldstein and P. Meystre, Phys. Rev. A {\bf 55},
2935 (1997).

\bibitem{Bashkin} E. P. Bashkin and A. V. Vagov, Phys. Rev. B {\bf 56}, 6207
(1997).

\bibitem{Graham} R. Graham and D. Walls, Phys. Rev. A {\bf 57}, 484 (1998).

\bibitem{Rapti1} Z. Rapti {\it et al.}, Phys. Lett. A {\bf 330}, 95 (2004).

\bibitem{Paraoanu} Gh. -S. Paraoanu, Phys. Rev. A {\bf 67}, 023607 (2003).

\bibitem{Timm} E. Timmermans, Phys. Rev. Lett. {\bf 81}, 5718 (1998).

\bibitem{AoChui} P. Ao and S.T. Chui, Phys. Rev. A {\bf 58}, 4836 (1998).

\bibitem{Kasa} K. Kasamatsu and M. Tsubota, Phys. Rev. Lett. {\bf 93}, 100402
(2004).

\bibitem{Burke} J. P. Burke {\it et al.}, Phys. Rev. A {\bf 55}, R2511 (1997).
\bibitem{Hall} D. S. Hall {\it et al.}, Phys. Rev. Lett. {\bf 81}, 1539
(1998).

\bibitem{Kivshar94} Y. S. Kivshar {\it et al.}, Phys. Rev. E {\bf 50}, 5020
(1994); M. Johansson and Y. S. Kivshar, Phys. Rev. Lett. {\bf 82}, 85 (1999).

\bibitem{Kevrekidis} P. G. Kevrekidis {\it et al.}, Phys. Rev. A {\bf 68},
035602 (2003).
\end{references}
\end{document}